\documentclass[aip,pop,reprint,frontmatterverbose]{revtex4-1}
\usepackage{graphicx}
\usepackage{ulem}
\usepackage{amssymb}
\usepackage{amsmath}
\usepackage{todonotes}
\usepackage{wasysym}
\usepackage{relsize}
\usepackage{bm}
\bibpunct{}{}{,}{s}{}{}
\usepackage{float}
\usepackage{epstopdf}

\begin{document}

\newcommand{\be}{\begin{equation}}
\newcommand{\ee}{\end{equation}}
\newcommand{\bea}{\begin{eqnarray}}
\newcommand{\eea}{\end{eqnarray}}
\newcommand{\bsp}{\begin{split}}
\newcommand{\esp}{\end{split}}
\newcommand{\Tbar}{{\bar{T}}}
\newcommand{\En}{{\cal E}}
\newcommand{\K}{{\cal K}}
\newcommand{\GC}{{\cal \tt G}}
\newcommand{\Lop}{{\cal L}}
\newcommand{\DB}[1]{\marginpar{\footnotesize DB: #1}}
\newcommand{\q}{\vec{q}}
\newcommand{\kt}{\tilde{k}}
\newcommand{\Lopn}{\tilde{\Lop}}
\newcommand{\noi}{\noindent}
\newcommand{\ovn}{\bar{n}}
\newcommand{\ovx}{\bar{x}}
\newcommand{\ovE}{\bar{E}}
\newcommand{\ovV}{\bar{V}}
\newcommand{\ovU}{\bar{U}}
\newcommand{\ovJ}{\bar{J}}
\newcommand{\calE}{{\cal E}}
\newcommand{\ovphi}{\bar{\phi}}
\newcommand{\zt}{\tilde{z}}
\newcommand{\rt}{\tilde{\rho}}
\newcommand{\tth}{\tilde{\theta}}
\newcommand{\nuv}{{\rm v}}
\newcommand{\ck}{{\cal K}}
\newcommand{\cc}{{\cal C}}
\newcommand{\ca}{{\cal A}}
\newcommand{\cb}{{\cal B}}
\newcommand{\cd}{{\cal D}}
\newcommand{\cg}{{\cal G}}
\newcommand{\ce}{{\cal E}}
\newcommand{\cn}{{\cal N}}
\newcommand{\fn}{{\small {\rm  FN}}}
\newcommand\norm[1]{\left\lVert#1\right\rVert}
\newcommand{\Afn}{A_{\text{FN}}}
\newcommand{\Bfn}{B_{\text{FN}}}
\newcommand{\mg}{\text{MG}}
\newcommand{\sh}{\text{SF}}
\newcommand{\tcc}{\text{cc}}
\newcommand{\tcm}{\text{cm}}
\newcommand{\tma}{\text{ma}}
\newcommand{\tmax}{\text{0}}

\title{Semi-analytical theory of emission and transport in a LAFE-based diode}

\author{Debabrata Biswas}\email{dbiswas@barc.gov.in}
\author{Rashbihari Rudra}
\author{Raghwendra Kumar}
\affiliation{
Bhabha Atomic Research Centre,
Mumbai 400 085, INDIA}
\affiliation{
  Homi Bhabha National Institute, Mumbai 400 094, INDIA}

\begin{abstract}
  A large area field emitter (LAFE) typically consists of several thousands of nanoscale emitting tips.
  These are difficult to simulate using purely numerical methods based on finite/boundary element or
  finite difference methods. We show here that a semi-analytically obtained electrostatic field allows
  tracking of field emitted electrons of a LAFE fairly accurately using the knowledge of only the LAFE
  geometry. Using a single and a 9-emitter configuration, the beam parameters calculated
  using this method are compared
  with the results of tracking using fields generated by COMSOL. The net emission current, energy conservation
  and the transverse trace-emittance are found to be reproduced with reasonable accuracy. 
\end{abstract}

\maketitle

\section{Introduction}
\label{sec:intro}

Large area field emitter (LAFE) based electron guns continue to be researched
due to their potential advantages such as fast switching, moderate to high beam brightness and a narrow energy distribution\cite{lewellen2005}. They are also being studied as an alternate
to photocathode injectors for use in free-electron
lasers \cite{leemann2007,jarvis2012}. Carbon nanotube based emitter
tips are particularly promising due to their inertness and stability,
and have been used in various devices such as portable
x-ray generators\cite{YuPark22,Hong18}.
While there has been considerable experimental progress, theoretical
studies and simulations have been few\cite{jensen2010,db_rudra18,db_fef,db_anode,rudra_db2019,db_rudra2020,db_hybrid2020,TA_DA_MC2020} due to several
challenges. A recent exception has been the
case of smooth vertically aligned emitters in a parallel plate diode geometry, where, it is now
possible to predict  local fields at each of the individual emitter apex and the
corresponding emission current to a reasonable accuracy using analytical
methods\cite{db_rudra18,db_fef,db_anode,rudra_db2019,db_rudra2020,db_hybrid2020}.
Note that there are typically
several thousands of high-aspect ratio emitting sites in a LAFE, often randomly placed, and simulating
these using a finite/boundary element software is a challenge in terms of the sheer amount
of resources required since there may be few or no symmetries that can be exploited for reducing the
size of the problem. It thus seems reasonable to pursue
a semi-analytical approach towards the simulation of a LAFE based diode.

A prerequisite for any such simulation requires an accurate estimate of the net current from
each emitter \cite{murphy,jensen_book,forbes2006,FD2007,db_dist,db_rr2019,db_rr2021,db_gamow,db_rk2019}
as well as the distribution of  electrons with respect to the position,
the  normal energy and the total energy\cite{db_dist,rk_gs_db2021,sarkar2021}.
The number of electrons together with their position and
momentum constitute the initial condition. This must be
supplemented with an accurate knowledge of the
electric field in the diode region, including the immediate neighbourhood of the sharp
emitting tip where the field changes rapidly\cite{rr_db2021}, in order to track the electrons as
they move from the cathode towards the anode. 

For a single or a few-emitter diode, there are numerous commercial software available
for tracking. A self-consistent solution requires more stringent particle-in-cell (PIC)
computations, which typically use finite-difference
algorithms for the ease of implementing charge conservation. In either case, the demand
on computational resources mount as the ratio between the height ($h$) and the
apex radius of curvature ($R_a$) increases\cite{min_domain} since even a small error in
the local field on the emitter-endcap gets exponentially amplified in the computation of the
field-emission current. Values of $h/R_a$ in the range $ 100 < h/R_a < 1000$
are not uncommon with $R_a$ as small as 5nm. An accurate computation of the local fields
for such high aspect-ratio emitters, requires a minimum mesh size smaller than $R_a/100$
and hence a very high mesh density. With a hundred such high-aspect ratio
emitters placed close together, the
task of finding just the field emission current from each distinct tip becomes an impossible task
for an average workstation. When particle data is added, the challenges increase.
Further, since the electrostatic field drops sharply
away from the apex within a few $R_a$, the grid size required for accurate tracking adds to the
computational resources. When thousands of such emitters are present, the task
becomes daunting even for a parallel machine.
It is thus necessary to explore analytical methods, both for emission and
tracking of the electrons in the diode region to supplement standard software used
for simulating macro-scale devices. If such an exercise is reasonably successful,
space-charge effects may subsequently be incorporated semi-analytically\cite{db_rr_gs2021}
in the modelling of LAFE-based diodes.

Analytical tracking methods have been used for a spherical diode
and for an emitter shape modelled by an equipotential of a `sphere mounted on a cone'
with another distant equipotential acting as an anode\cite{everhart67,everhart73,everhart74}.
The availability of analytical
solutions for such geometries makes them particularly attractive and the model has been
used to track and evaluate beam parameters.

Our interest in this work lies in a parallel plate diode geometry with identical vertically
aligned emitters mounted on the cathode plate, and the anode placed in close proximity
to the tip(s) such as in a gated diode. We shall employ semi-analytical approximate
methods, based on the non-linear line charge model\cite{nlcm2016,mesa,pogo2009,harris15,harris16}
(NLCM), to estimate the current from each emitter
and subsequently track the emitted electrons, for a single emitter initially and
eventually for several emitters. The effectiveness of the semi-analytical
method can be gauged from a  comparison  with COMSOL generated electrostatic field data, both
for the computation of the field-emission current and the tracking of electrons in the diode
region. The emitted electrons in both the semi-analytical and COMSOL routes follow
the same distribution but may differ in number due to the difference in emission
current which in turn is related to the apex field. Note that the semi-analytical apex
field and the COMSOL generated apex field are bound to differ even if slightly,
leading to errors in the emission current that may be considerably amplified.
We shall ignore space charge effects in the following
and focus merely on the net emission current, energy-conservation and the average transverse emittance.
The issue of emitters having non-identical height or apex radius of curvature
is yet another complication that will not be addressed in the present communication.

Recent progress in using
an NLCM-based hybrid model\cite{db_hybrid2020} shows that effects such as shielding, anode-proximity
and inverse-shielding mediated
through the anode, can be captured fairly accurately using purely geometric quantities
such as the height ($h$), apex radius of curvature ($R_a$), the coordinates of each emitter ($x_i,y_i$)
on the cathode plane ($z = 0$) and the anode-cathode distance ($D$). The estimate is good
if the emitters are not too close although even an error as small as 5\% in the apex field can translate
into 25-50\% change in the emitted current. Deviations of this order from the true value are par
for the course, keeping in mind that small experimental uncertainties 
in estimating the height and radius of curvature can result in even larger changes
in current. Having estimated the current, the emitted electrons from each emission site needs to be
transported across the diode using fields generated using the NLCM based hybrid model.
The parameters for judging the efficacy of the semi-analytical transport
include energy conservation and transverse emittance at the anode\cite{reiser,floettmann2003,brau}.

In the following sections, we shall first review briefly the semi-analytical hybrid model
and follow it up with simulations for a single hemiellipsoidal emitter
and a hemiellipsoid on a cylindrical post (HECP) to illustrate the method and
estimate its efficacy. We shall then use a 9 emitter cathode
and compare the results with COMSOL generated electrostatic fields.
Summary and discussions form the concluding section.

\section{The NLCM based hybrid model}
\label{sec:model}

\begin{figure}[hbt]
  \begin{center}
    \vskip -1.75cm
\hspace*{.250cm}\includegraphics[scale=0.33,angle=0]{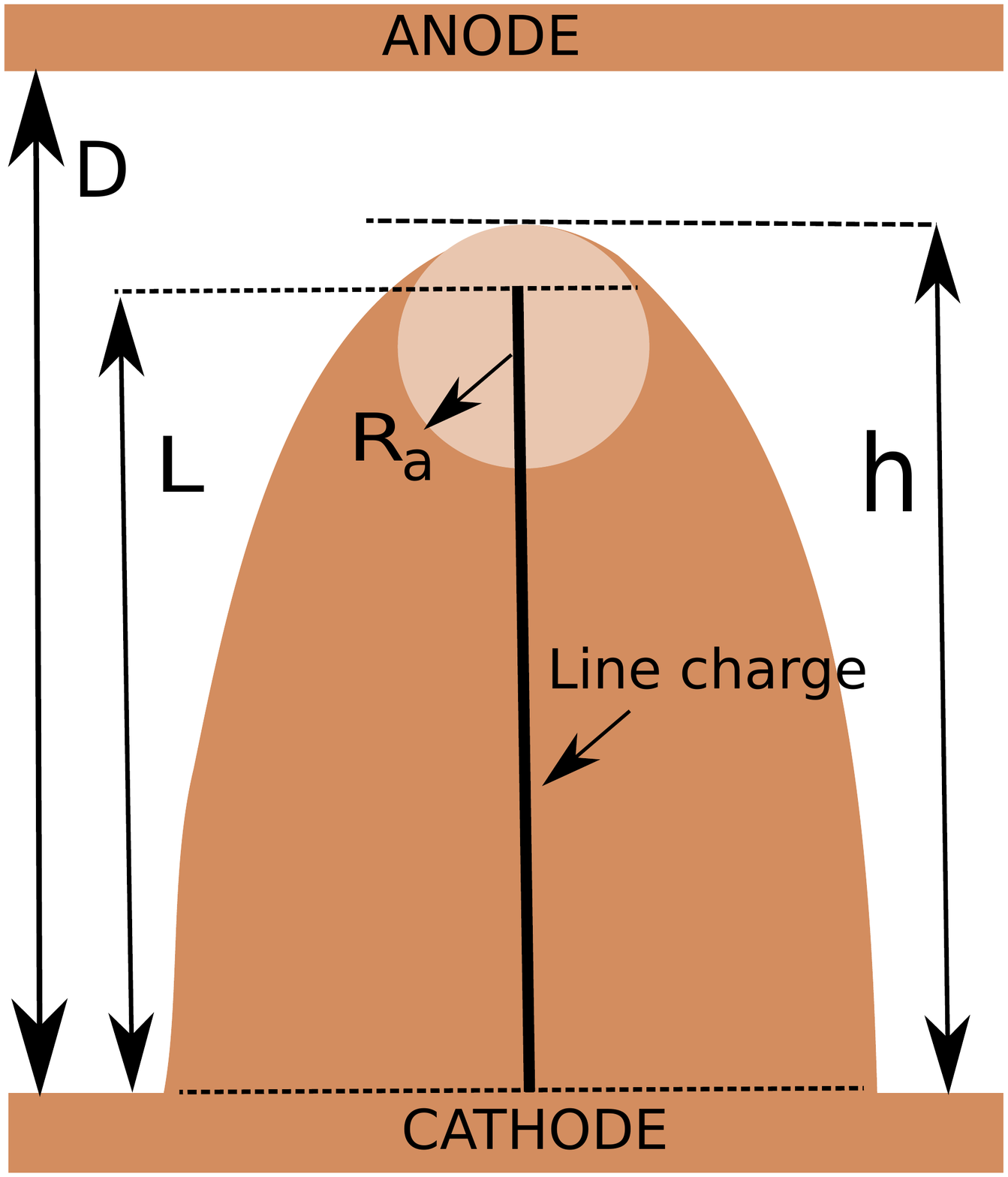}
\vskip -0.2 cm
\caption{The emitter can be represented as an equipotential generated by
  a linecharge and an applied electric field $-E_0 \hat{z}$.}
\label{fig:linecharge}
\end{center}
\end{figure}

Consider an emitter of height $h$ and apex radius of curvature $R_a$ as shown
in Fig.~\ref{fig:linecharge}, mounted on
a grounded cathode plate at $z = 0$ with the anode at $z = D$ having a potential $V_g$.
If the emitter is a hemi-ellipsoid and the anode far away, the system is identical
to a line charge placed vertically on the cathode plane extending from $z = 0$
to $z = L = h - R_a/2$ having a linear charge density $\lambda$, and its image
extending from $z = 0$ to $z = -L$ having a linear charge density $-\lambda$.
If the field far away from the tip is expressed as $E_0 = V_g/D$, it can be shown that

\be
\lambda = - \frac{4\pi\epsilon_0 E_0}{\ln[(h+L)/(h-L)] - 2L/h}. \label{eq:hemi}
\ee

\noi
For a sharp emitter where $h/R_a >> 1$, 
$\lambda \approx -4\pi\epsilon_0 E_0/(\ln(4h/R_a) - 2)$. The line charge is in effect the
projection of the induced surface charge  on the hemiellipsoid
surface on to the axis of symmetry\cite{nlcm2016}.

The potential at any point ($\rho,z$) due to a vertical line charge placed on a grounded conducting
plane can be expressed as

\be
\begin{split}
V(\rho,z) = & \frac{1}{4\pi\epsilon_0}\Big[ \int_0^L \frac{\Lambda(s)}{\big[\rho^2 + (z - s)^2\big]^{1/2}} ds ~
  - \\
  &  \int_0^L \frac{\Lambda(s)}{\big[\rho^2 + (z + s)^2\big]^{1/2}} ds \Big] + E_0 z \label{eq:pot}
\end{split}
\ee

\noi
where $\Lambda(s) = \lambda s$ and $L$ is the extent of the line charge distribution.
The equipotential curve $V = 0$ generated by the line charge
and its image together with the macroscopic field $E_0$, has the shape of a
hemiellipsoid.

As the anode is brought near, this description becomes somewhat inaccurate since
a linear line charge density $\lambda(s)$ cannot strictly generate a hemiellipsoidal
equipotential surface when the effects of the anode are incorporated.
Nevertheless, a reasonably accurate description can be obtained by persisting
with a linear line charge density, modified as\cite{db_anode}

\be
\lambda \approx - \frac{4\pi\epsilon_0 E_0}{\ln[4h/R_a] - 2 - \alpha_A} \label{eq:hemianode}
\ee

\noi
where $\alpha_A$ represents the infinitely many images of the original line-charge pair
formed due to reflection by the anode and cathode planes and is expressed as\cite{db_anode}

\be
\begin{split}
  \alpha_A & = \frac{2}{h}\sum_{n=1}^\infty \left[(2nD-h)\tanh^{-1}\frac{L}{2nD-h} \right. \\
  & - \left. (2nD+h)\tanh^{-1}\frac{L}{2nD+h}\right] \label{eq:alpA}.
  \end{split}
\ee

If we are dealing with a collection of emitters placed at ($x_i,y_i$) on the cathode
plane, their collective effect can be expressed by modifying the line charge density
of the $i^{th}$ emitter as\cite{db_rudra2020,db_hybrid2020}

\be
\lambda_i \approx - \frac{4\pi\epsilon_0 E_0}{\ln[4h/R_a] - 2 - \alpha_A + \alpha_{S_i} -\alpha_{SA_i}} \label{eq:hemifull}
\ee

\noi
where $\alpha_{S_i}$ is the shielding contribution on the $i^{th}$ emitter due to all the other
emitters while $\alpha_{SA_i}$ is the inverse shielding effect mediated through the anode.
They are expressed respectively as\cite{db_rudra2020,db_hybrid2020}

\be
\begin{split}
  \alpha_{S_i} \simeq & \sum_{j\neq i}^{N}  \Bigg[ \frac{1}{h}\sqrt{\rho_{ij}^2 + (h - L)^2} - \frac{1}{h}\sqrt{\rho_{ij}^2 + (h + L)^2} \\
    & + \ln\Bigg(\frac{\sqrt{\rho_{ij}^2 + (h + L)^2} + h + L}{\sqrt{\rho_{ij}^2 + (h - L)^2} + h - L}\Bigg) \Bigg] \label{eq:alpS}
\end{split}
\ee

\noi
while

\be
\begin{split}
  \alpha_{{SA}_i} \simeq & \sum_{n=1}^\infty \sum_{j\neq i}^{N}  \Bigg[
    \frac{\cd_{mm}}{h} - \frac{\cd_{mp}}{h} - \frac{\cd_{pm}}{h} + \frac{\cd_{pp}}{h} \\
    & + \frac{2nD - h}{h} \ln\Big(\frac{\cd_{mp} + 2nD - h + L}{\cd_{mm} + 2nD - h - L}\Big) \\
    & - \frac{2nD + h}{h} \ln\Big(\frac{\cd_{pp} + 2nD + h + L}{\cd_{pm} + 2nD + h - L}\Big) \Bigg] \label{eq:alpSA}
\end{split}
\ee

\noi
with

\bea
\cd_{mm} & = & \sqrt{\rho_{ij}^2 + (2nD - h - L)^2} \nonumber \\
\cd_{mp} & = & \sqrt{\rho_{ij}^2 + (2nD - h + L)^2} \nonumber \\
\cd_{pm} & = & \sqrt{\rho_{ij}^2 + (2nD + h - L)^2} \nonumber \\
\cd_{pp} & = & \sqrt{\rho_{ij}^2 + (2nD + h + L)^2}. \nonumber
\eea

\noi
In the above $\rho_{ij} = \sqrt{x_{ij}^2 + y_{ij}^2}$ is the distance between the $i^{\text{th}}$ and $j^{\text{th}}$ emitter on the cathode plane ($\text{XY}$).

\subsection{The field at the apex}
\label{subsec:apex}

With these inputs, the field at the apex of the $i^{th}$ emitter can be expressed as

\be
E_a^{(i)} \simeq E_0 ~\frac{2h/R_a}{\ln(4h/R_a) - 2 - \alpha_A + \alpha_{S_i} -\alpha_{SA_i}} = \gamma_a^{(i)} E_0 \label{eq:Ea}
\ee

\noi
where $\gamma_a^{(i)}$ is referred to as the apex field enhancement factor of the $i^{th}$ emitter. Thus, if the co-ordinates
of a collection of $N$ emitters ($N$ can be large) are known, the approximate field at the apex of each emitter
can be evaluated fairly easily. Coupled with the generalized cosine law of local field variation near the emitter
apex\cite{db_ultram,db_physE}

\be
E_l(\tth) = E_a \frac{z/h}{\sqrt{(z/h)^2 + (\rho/R_a)^2}} = E_a \cos\tth, \label{eq:costheta}
\ee

\noi
the current from each of them can be evaluated simply from a knowledge of the apex field $E_a^{(i)}$ and
the radius of curvature $R_a$.

As stated earlier, in reality a linear line charge does not have a hemiellipsoidal equipotential when the anode
and all the other emitters are thrown in. Besides, if the shape of the emitter itself is different (such as
a rounded cone or a cylindrical post), the line charge density is required to be nonlinear to begin with even if the
emitter is isolated. The use of a nonlinear line charge leads to expressions of the form\cite{db_anode}

\be
\gamma_a \approx \frac{(1 - \cc_0)~2h/R_a}{(1 - \cc_1) \ln[4h/R_a] - (1 - \cc_2) 2 - (1 - \cc_3) \alpha_A} \label{eq:hemianode1}
\ee

\noi
when an isolated emitter is in close proximity to the anode. Here $( 1 - \cc_i)$, are
corrections due to the nonlinear line charge density but are a-priori unknown.
The number of such unknowns can be
reduced and $\gamma_a$ simply expressed as

\be
\gamma_a \approx  \frac{2h/R_a}{\alpha_1 \ln[4h/R_a] - \alpha_2} \label{eq:hemianode2}
\ee

\noi
where $\alpha_1$ and $\alpha_2$ can be determined by re-writing Eq.~(\ref{eq:hemianode2}) as\cite{sarkar_db2019}

\be
\frac{2h/R_a}{\gamma_a} = {\alpha_1 \ln[4h/R_a] - \alpha_2}  \label{eq:finda12}
\ee

\noi
and plotting $2h/(R_a \gamma_a)$ vs $\ln(4h/R_a)$. The quantities  $h/R_a$ can
be varied by keeping $h$ and $D$ constant while changing $R_a$ locally around the desired value.
For each such value of $h/R_a$, the apex field enhancement factor $\gamma_a = E_a/E_0$
can be determined using a suitable software
such as COMSOL. A straight line fit to the data
has $\alpha_1$ as the approximate slope and $\alpha_2$ as the intercept.

Once $\alpha_1$ and $\alpha_2$ are known, the apex field can be determined using

\be
E_a^{(i)} \simeq E_0 ~\frac{2h/R_a}{\alpha_1 \ln(4h/R_a) - \alpha_2 + \alpha_{S_i} -\alpha_{SA_i}}  \label{eq:Ea1}.
\ee

\noi
Note that the anode correction $\alpha_A$ has been incorporated in $\alpha_2$ and can be dropped altogether. Also, we have ignored the corrections due to the nonlinear line charge while expressing
the contributions of the other emitters. Thus, Eq.~(\ref{eq:Ea1}) is not expected to be too
accurate when the emitters are closely packed but is a fair approximation so long as the
mean spacing is greater than the height of the emitters. Since the approximate apex
field is known, the current from an $N$-emitter system can thus be determined semi-analytically.

\subsection{The field in the diode region}

The field at the apex of the sharp emitters is much larger than $E_0 = V_g/D$. In its immediate vicinity
outside the emitter surface, the field remains high. Determining this variation accurately as
the point moves away from the apex requires enormous computational resources.
The line charge model provides an alternate approximate economical means to determine the field
components in the diode region since $\lambda$ in Eq.~(\ref{eq:hemifull}) does approximately
take into account the anode and other emitters in the neighbourhood. We shall continue
with the use of $\lambda$ even when the line charge density is expected to be nonlinear
with the caveat that $\alpha_1$ and $\alpha_2$ are determined numerically for a single
emitter as described in section \ref{subsec:apex} (in particular Eq.~\ref{eq:finda12}).
This implies that the effect of nonlinearity at the single-emitter level is incorporated
but it is ignored while expressing the contribution of other emitters.

Within this approximate model, the $\hat{\rho}$ component of the electric field,
-$\partial V/\partial \rho$ at the point ($\rho,z$) can be evaluated to yield

\be
\begin{split}
E_\rho = \frac{\lambda}{4\pi\epsilon_0} \frac{1}{\rho} \Bigg[ \frac{\rho^2 + z(z+ L)}{\sqrt{\rho^2 + (z + L)^2}} - &
  \frac{\rho^2 + z(z - L)}{\sqrt{\rho^2 + (z - L)^2}} \Bigg]
\end{split}
\ee

\noi
while the $\hat{z}$ component of the field -$\partial V/\partial z$ is

\be
\begin{split}
E_z = & \frac{\lambda}{4\pi\epsilon_0}  \Bigg[ \frac{L}{\sqrt{\rho^2 + (z + L)^2}} + \frac{L}{\sqrt{\rho^2 + (z - L)^2}} ~+  \\
&  \ln\Big\{ \frac{\sqrt{\rho^2 + (z+L)^2} - (z+L)}{\sqrt{\rho^2 + (z-L)^2} - (z-L)} \Big\}\Bigg] - E_0 . \label{eq:Ez0}
\end{split}
\ee

\noi
For small values of $\rho$ (i.e. close to the axis of symmetry),

\be
E_\rho \simeq \frac{\lambda \rho}{4\pi\epsilon_0} \Bigg[ \frac{2z^2 L}{(z^2 - L^2)^2} - \frac{2L}{(z^2 - L^2)} \Bigg]
\ee

\noi
while

\be
E_z =  \frac{\lambda}{4\pi\epsilon_0}  \Bigg[ \frac{L}{\sqrt{\rho^2 + (z + L)^2}} + \frac{L}{\sqrt{\rho^2 + (z - L)^2}} \Bigg] - E_0. \label{eq:Ez0}
\ee

\noi
We shall first test how these fields fare for a single emitter before applying them to a collection of emitters.

\section{Numerical tests}
\label{sec:single}

\subsection{Single emitters}

We shall first consider two examples of single field emitters. The first is a hemiellipsoid (HE) with $h/R_a = 300$ and
$R_a = 500$nm. The other is a hemiellipsoid on a cylindrical post (HECP), again with  $h/R_a = 300$
with $R_a = 500$nm. The hemiellipsoid endcap has a height $h_e = 5R_a$
while the cylindrical post has height $h - 5R_a$ and radius $\sqrt{h_e R_a}/2$.
The anode in both cases is placed at a distance $20R_a$ from the tip so that the
anode-cathode plate distance is $D = 160\mu$m and $h = 150\mu$m. The hemiellipsoid with the anode far
away (e.g. $D > 5h$) is analytically solvable with $\alpha_1 = 1$ and $\alpha_2 = 2$ for the
values of $h$ and $R_a$ chosen. It is expected that tracking would be accurate in such a case.
With the anode placed close to the tip, a simple closed form analytical solution does not exist
and this scenario presents the first non-trivial test of the method outlined in section \ref{sec:model}.
The HECP emitter on the other hand has no analytical solution even when the anode is far away and
is a typical case where the line charge density is nonlinear. The two examples thus present
non-trivial test cases for the methodology described in section \ref{sec:model}.

In order to use the semi-analytical model outlined in section \ref{sec:model} and test its
accuracy, we shall first determine the electrostatic field using  the finite-element based
multiphysics simulation software, COMSOL. Since the problem eventually involves multiple emitters,
a 3-dimensional depiction of the hemiellipsoid and HECP emitters is used. These are placed
in a parallel plate geometry  with the bottom (cathode)-plate
having $V = 0$ while the anode is another equipotential with $V = 4800$V. The computational
domain is a cuboid with the other 4 faces placed at a distance $10h$ from the centre of the
box, each having Neumann boundary condition. With $h/R_a = 300$ and $R_a = 500$nm, the minimum
domain size is chosen to be less than or equal to $R_a/100$. Convergence is
ensured in each case by adjusting the
mesh parameters.

\begin{figure}[hbt]
  \begin{center}
    \vskip -0.9cm
\hspace*{-1.050cm}\includegraphics[scale=0.37,angle=0]{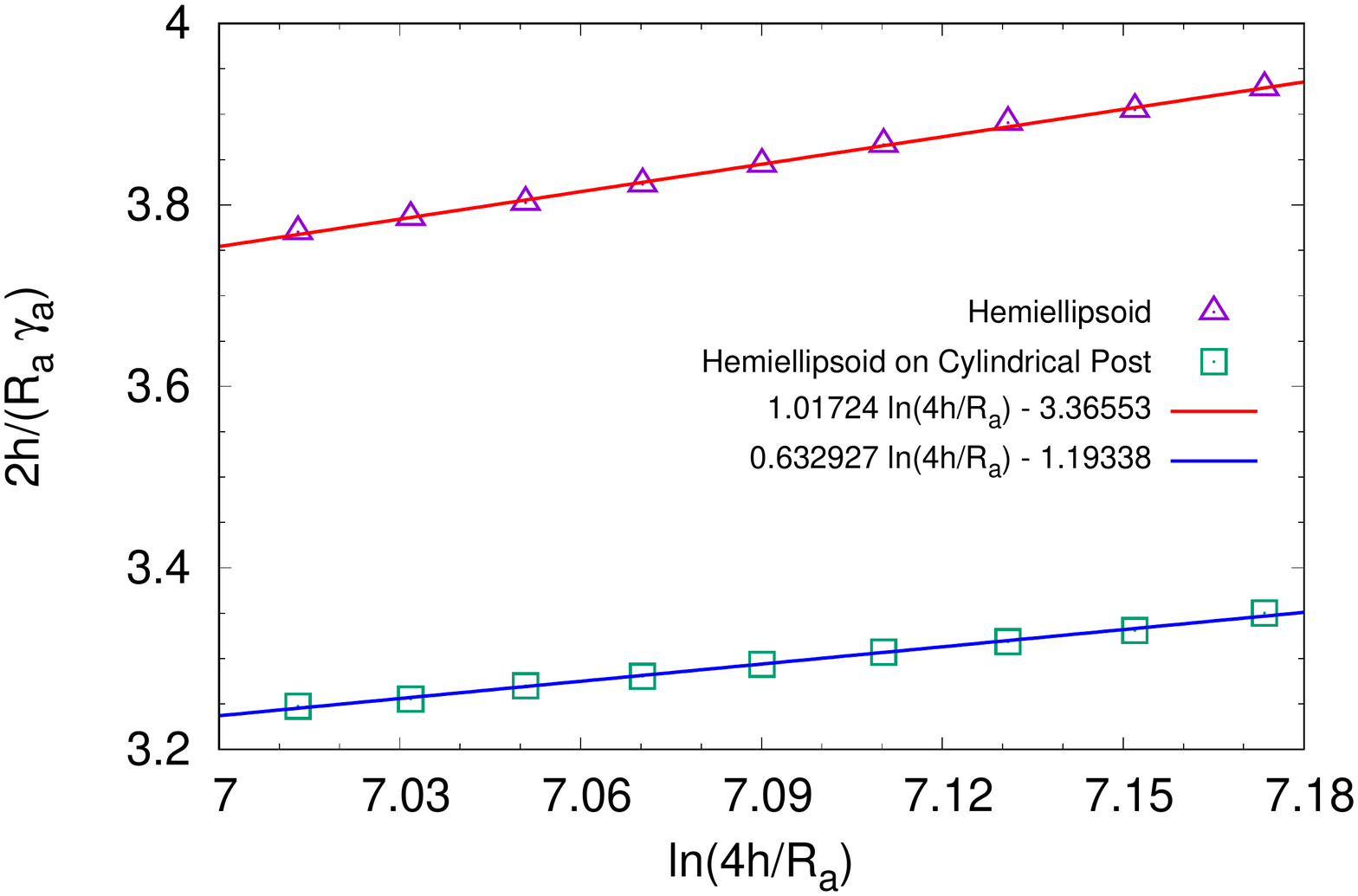}
\vskip -0.5 cm
\caption{The values of $(2h/R_a)/\gamma$ are shown plotted against $\ln(4h/R_a)$. They appear
  to be on a straight line. The best fit is also shown. The slope and the intercept are
estimates of $\alpha_1$ and $\alpha_2$ respectively.}
\label{fig:alp12}
\end{center}
\end{figure}

We shall consider a single emitter placed at the centre of the cathode plate.
Fig.~\ref{fig:alp12} shows the distribution of $(2h/R_a)/\gamma_a$ values plotted against $\ln(4h/R_a)$.
The apex field enhancement factor $\gamma_a$ is obtained using COMSOL for both the Hemiellipsoid (HE)
and hemiellipsoid on a cylindrical post (HECP). Notice that both sets of points lie approximately on
a straight line. The best fits are also shown with the slope representing $\alpha_1$ and the
intercept $\alpha_2$. Thus, for the hemiellipsoid emitter with the anode at $D = h + 20R_a$,
$\alpha_1 = 1.01724$ while $\alpha_2 = 3.36553$. Notice that the anode-at-infinity should have
$\alpha_1 = 1$ and $\alpha_2 = 2$. The increase in $\alpha_2$ is due to the anode proximity effect.

In case of the hemiellipsoid on a cylindrical post (HECP), the best fit value of the pair
is $\alpha_1 = 0.632927$ and $\alpha_2 = 1.19338$. Since, the case corresponds to a nonlinear
line charge and $\cc_0$,$\cc_1$,$\cc_3$ and $\cc_4$ are unknown, we shall use the best
fit values $\alpha_1$ and $\alpha_2$ to represent an HECP with the anode placed at $D = h + 20R_a$.

\begin{figure}[hbt]
  \begin{center}
    \vskip -0.9cm
\hspace*{-1.050cm}\includegraphics[scale=0.37,angle=0]{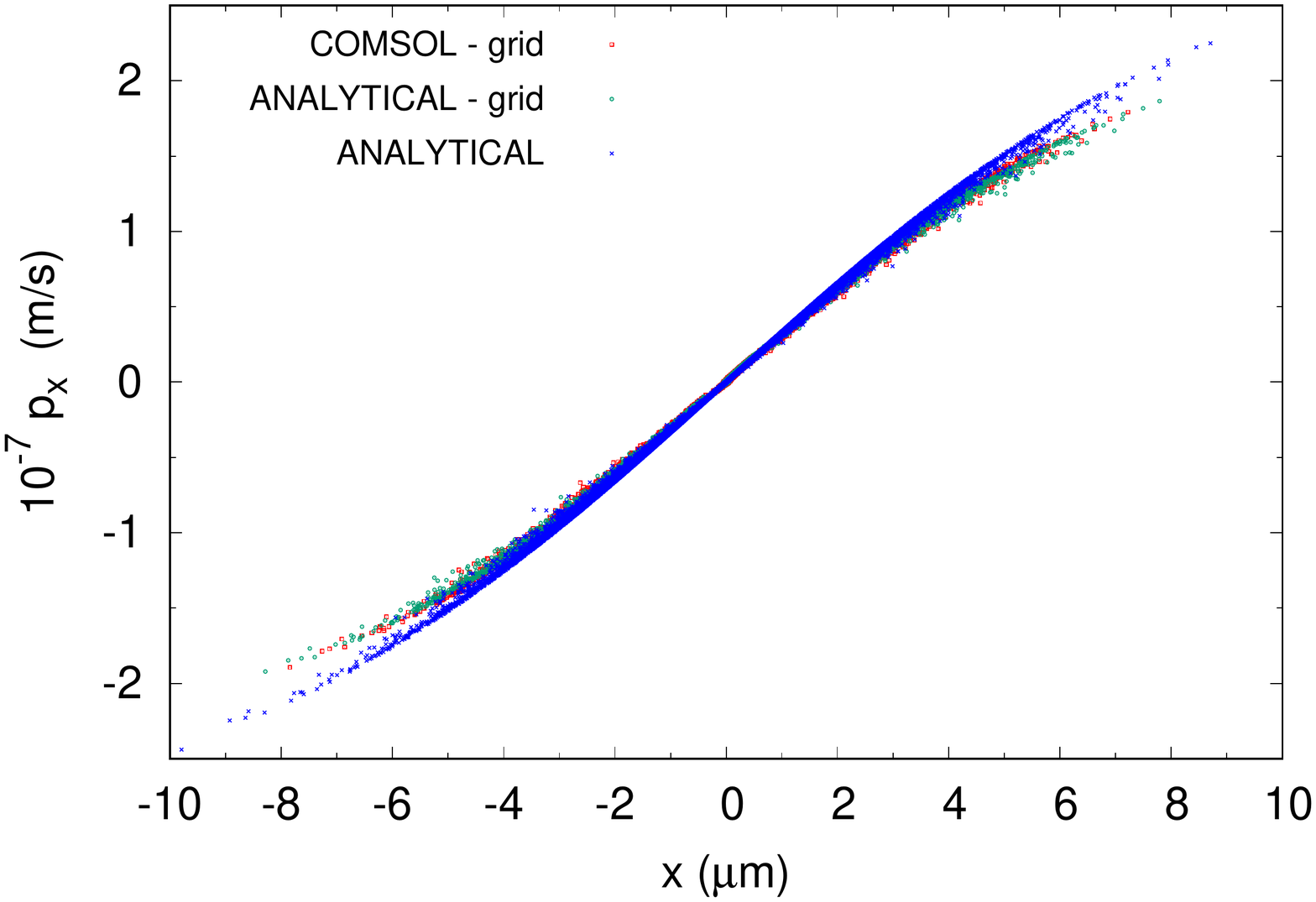}
\vskip -0.95 cm
\caption{An $x-p_x$ plot at the anode located at $z = 160\mu$m. The $y-p_y$ plot looks similar. The
emitter is a hemiellipsoid with apex radius $R_a = 500$nm and height $h = 150\mu$m.}
\label{fig:xpx}
\end{center}
\end{figure}

Having determined $\alpha_1$ and $\alpha_2$, the apex field $E_a$ and the
line charge density $\lambda$ can be obtained using the value of the macroscopic
field $E_0 = 3 \times 10^7$ V/m. These can in turn be used to generate the
$N$ particle positions on the endcap using the distribution function\cite{db_dist,rk_gs_db2021}

\be
f({\tilde \theta}) \approx 2 \pi R_a^2 {\frac{\sin{\tilde \theta}}{\cos^2 {\tilde \theta}}} J_{\mg}({\tilde \theta})
\label{Eq:dist}
\ee

\noi
where $\tth$ is given by Eq.~(\ref{eq:costheta}) and $J_{MG}$ is the Murphy-Good current density

\begin{equation}
  J_{\mg}({\tth}) = \frac{1}{t_F^2}\frac{A_{FN}}{\phi} (E_l({\tth}))^2 \exp\left(-{B_{FN}}{v_F} {\phi}^{3/2}/E_l({\tth})\right).
    \label{eq:MG}
\end{equation}

\noi
Here 
$A_\fn~\simeq~1.541434~{\rm \mu A~eV~V}^{-2}$,
$B_\fn~\simeq 6.830890~{\rm eV}^{-3/2}~{\rm V~nm}^{-1}$ are the conventional Fowler-Nordheim constants
while $v_F \simeq 1 - f + (f/6) \ln f$ and $t_F \simeq 1 + f/9 - (f/18)\ln f$ are corrections due to the image
charge potential with 
$f  \equiv  c_S^2~E_l({\tth})/\phi^2$ where $c_S$ is the Schottky constant and
$c_S^2 = 1.439965~ \text{eV}^2 \text{V}^{-1} \text{nm}$. $E_l({\tth})$ refers to the local electric field on the emitter surface, while $\phi$ is the work function of the emitter material which we shall consider to be 4.5eV.
In all the studies presented here, $N = 20000$.

As mentioned in Ref.~[\onlinecite{db_dist},\onlinecite{sarkar2021},\onlinecite{rk_gs_db2021}], first ${\tilde \theta}$ is sampled from the distribution described above. The value of the normal energy $\mathcal{E_N}$ and total energy $\mathcal{E_T}$ are thereafter obtained using the conditional distributions $f(\mathcal{E_N}|{\tilde \theta})$ and $f(\mathcal{E_T}|\mathcal{E_N},{\tilde \theta})$ respectively. These distributions can be arrived at using the joint distribution  $f({\tilde \theta},\mathcal{E_N},\mathcal{E_T})$ \cite{db_dist}. Once the angle $\tth$, normal energy and total energy are calculated, velocity components in local co-ordinate system are obtained using suitable transformations.

There are three kinds of tracking that we shall focus on. The {\it first} uses the analytical fields
directly to integrate the trajectory from the emitter surface to the anode. The {\it second} makes use
of analytical field computed on a grid while the {\it third} uses the
COMSOL field on a grid\cite{comsol_use,rudra_db2019}. As the
data typically available in a simulation is on a grid (with appropriate weighting schemes 
to find the point data), the analytical and COMSOL fields are best compared by considering
identical grids. The grid chosen here consists of
$64 \times 64 \times 256$ points with 64 each in the X and Y directions extending
from [$-3.25\mu$, $3.25\mu$m]
and 256 points along the Z-direction from $149\mu$m to $161\mu$m. The anode is placed at $160\mu$m
and is assumed to have a thickness of $1\mu$m. The position and momentum of the
electrons are recorded when they reach the anode and these are subsequently used for further
analysis.

\begin{figure}[hbt]
  \begin{center}
\hspace*{-1.050cm}\includegraphics[scale=0.37,angle=0]{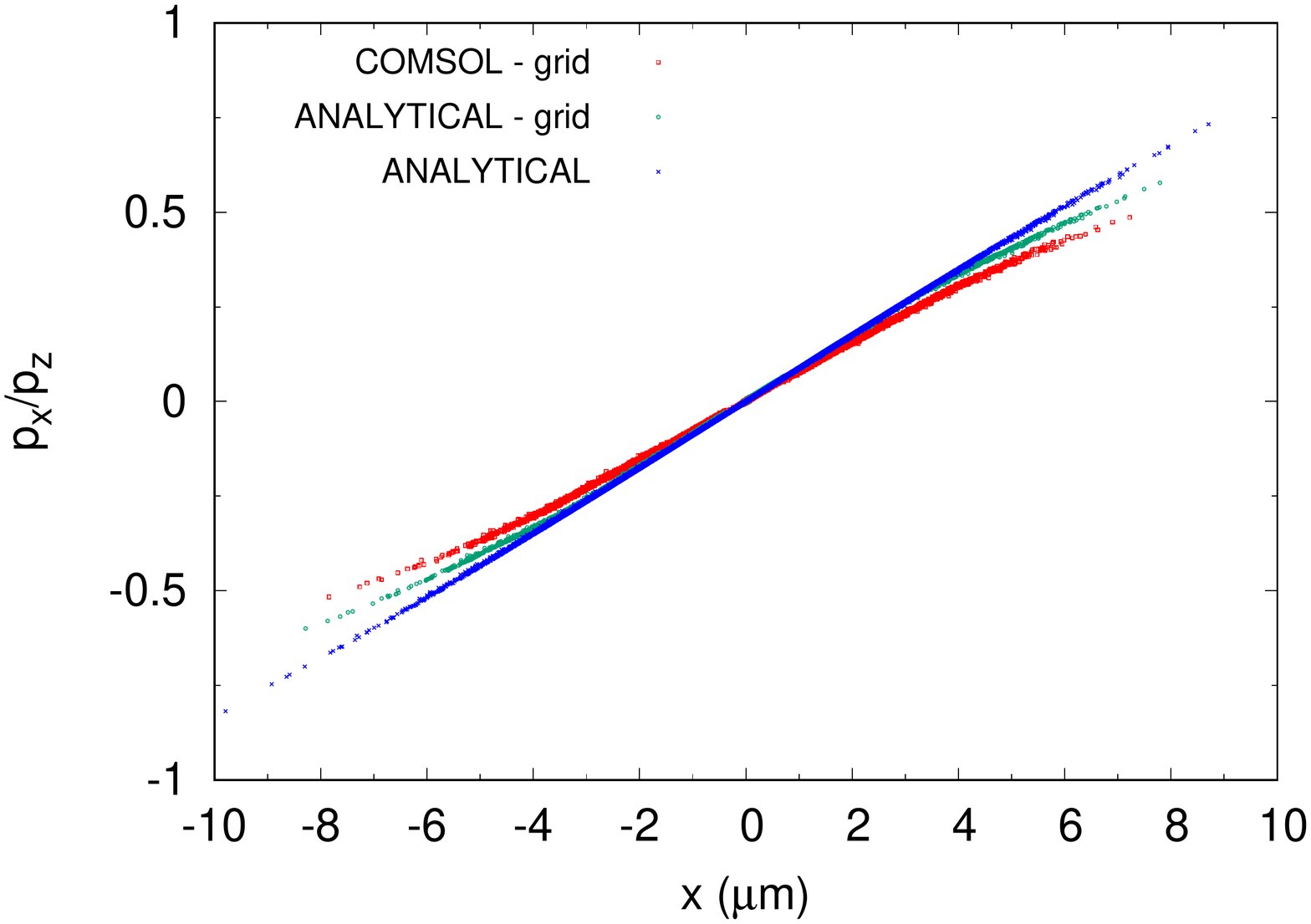}
\vskip -0.35 cm
\caption{A trace-space $x-x'$ plot at $z = 160\mu$m where $x' = p_x/p_z$.The
emitter is a hemiellipsoid of apex radius $R_a = 500$nm.}
\label{fig:xxprime}
\end{center}
\end{figure}

\begin{figure}[hbt]
  \begin{center}
    \vskip -0.9cm
\hspace*{-1.050cm}\includegraphics[scale=0.37,angle=0]{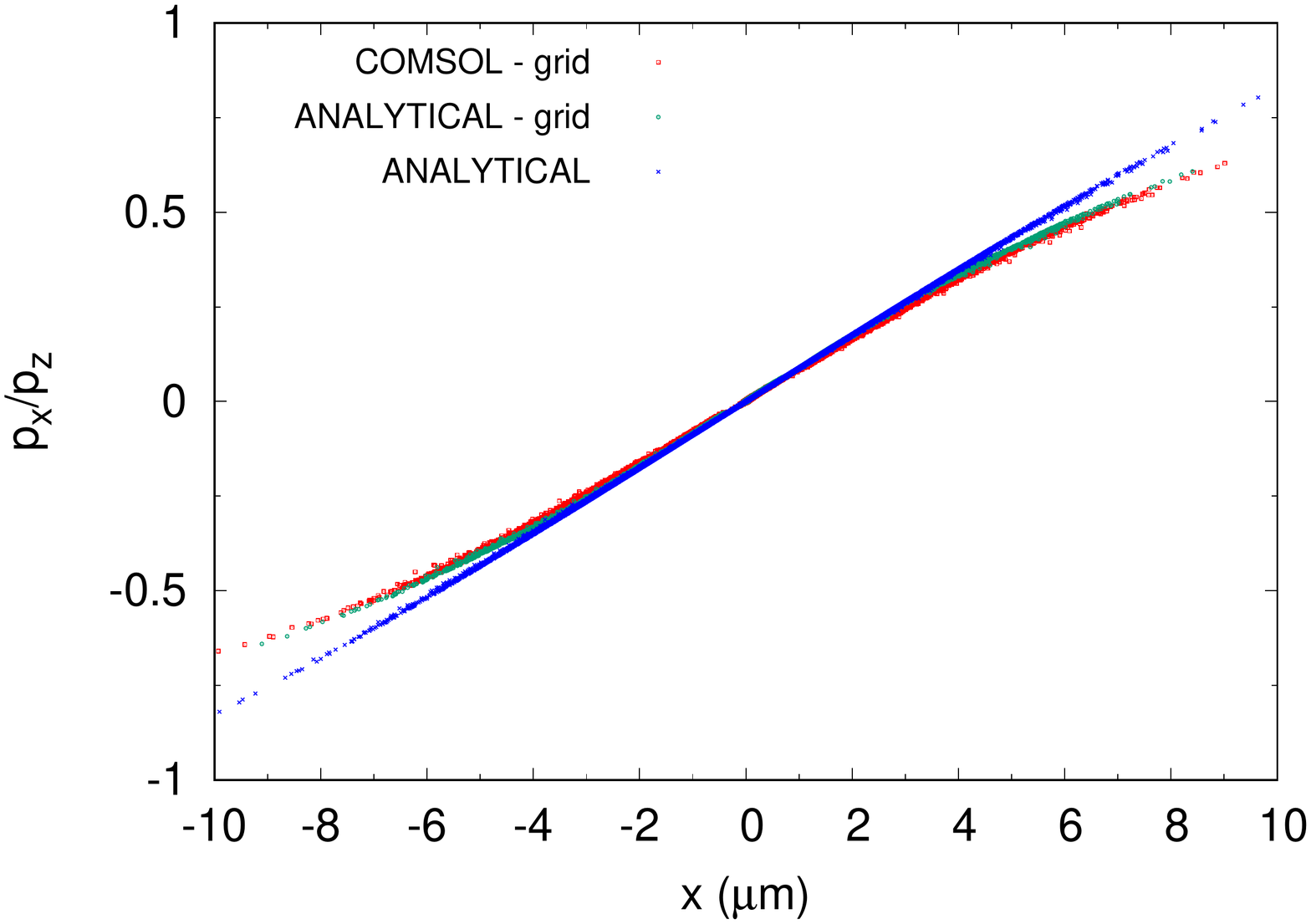}
\vskip -0.5 cm
\caption{A trace-space plot at $z = 160\mu$m for the HECP emitter.}
\label{fig:xxprime_hecp}
\end{center}
\end{figure}

A phase space $x-p_x$ plot for the hemiellipsoid emitter is shown in Fig.~\ref{fig:xpx}.
Clearly, the grid data for the analytical and COMSOL fields are close to each
other while non-grid analytical result shows some deviation at the extremities.
It is common however to study the trace-space plots where $x' = p_x/p_z$ and $y' = p_y/p_z$
are used instead. A plot of $x-x'$ is shown in Fig.~\ref{fig:xxprime} for the
hemiellipsoid emitter. The difference between the analytical and COMSOL grid-data is
more evident now. A similar plot for the HECP emitter is shown in Fig.~\ref{fig:xxprime_hecp}

A quantitative estimate of the deviation can be obtained from the rms trace-emittance
defined as\cite{floettmann2003}

\be
\epsilon_{\text{Tr,rms}} = \sqrt{<x^2> <x'^2> - <xx'>^2} = \epsilon_x
\ee

\noi
where 

\bea
<x^2> & = & \frac{\sum_i x_i^2}{N} - \left( \frac{\sum_i x_i}{N} \right)^2 \\
<x'^2> & = & \frac{\sum_i {x'}_i^2}{N} - \left( \frac{\sum_i {x'}_i}{N} \right)^2 \\
<xx'>^2 & = & \frac{\sum_i x_i{x'}_i}{N} -  \frac{\sum {x'}_i \sum x_i}{N^2}
\eea

\noi
Table \ref{tab:symbols} which summarizes
the results for the single-emitters considered. The quantities of interest are (i) the emitter current
(ii) the apex field  (iii)$ \int E.dl = \int_h^D E_z dz$ along the axis of symmetry and
finally $\epsilon_x$.

It is clear from Table \ref{tab:symbols} that the values of $\alpha_1$ and $\alpha_2$, obtained through a
fit, allow accurate determination of the apex field and hence the emission current. The values of
$\int E.dl$ clearly show that while COMSOL evaluates the diode field accurately, the analytical field
has errors which reflects both in the grid and non-grid evaluations. For the HE emitter, the analytical field
under-predicts $\int E.dl$ while for the HECP emitter, it over-predicts. The error ranges from 2.6\% for HE 
to 12.6\% for HECP for non-grid data, while the grid-data has errors of 8.9\% for HE and 6.3\% for HECP.
The transverse emittance values are reasonably close for the COMSOL and analytical grid data, supportive
of Figs. \ref{fig:xxprime} and \ref{fig:xxprime_hecp}.

\begin{table}[htb]
  \begin{center}
    \vskip -0.25 in
   \caption{Single-emitter characteristics. The symbols are HE (Hemiellipsoid), HECP (Hemiellipsoid on Cylindrical Post), ANLYTL (Analytical), Y (Yes) and N (No), I (Current). For both emitters, the total height $h = 150\mu$m and the anode-cathode distance $D = 160\mu$m with apex radius of curvature $R_a = 500$nm. The correct $\int E.dl$ value is 4800V. }
   \vskip 0.05 in
    \label{tab:symbols}
    \begin{tabular}{|l|l|c|l|c|c|c|}
      \hline
      Emitter & Method & Grid & $~~~E_a$  & I  & $ \int E.dl$   &  $\epsilon_x$  \\ 
      & & & (V/nm) & (mA) & (V) & (nm) \\  \hline \hline
      HE & ANLYTL & N & 4.7081 & 0.18961 & 4674  & 3.748 \\ \hline
      HE & ANLYTL & Y & 4.7081 & 0.18961 & 4371 & 9.958 \\ \hline
      HE & COMSOL & Y & 4.6821 & 0.17347 & 4799 & 5.985 \\ \hline
      HECP & ANLYTL & N & 5.4899 & 1.91510 & 5405 & 5.203 \\ \hline
      HECP & ANLYTL & Y & 5.4899 & 1.92510 & 5101 & 13.921 \\ \hline
      HECP & COMSOL & Y & 5.4648 & 1.79525 & 4796 & 12.436 \\ \hline
    \end{tabular}
\end{center}
\end{table}

In summary, single emitter predictions for the current and transverse-emittance are reasonably accurate
for single emitters.

\subsection{A 9-emitter array}

\begin{figure}[hbt]
  \begin{center}
    \vskip -2.750cm
\hspace*{-.050cm}\includegraphics[scale=0.5,angle=0]{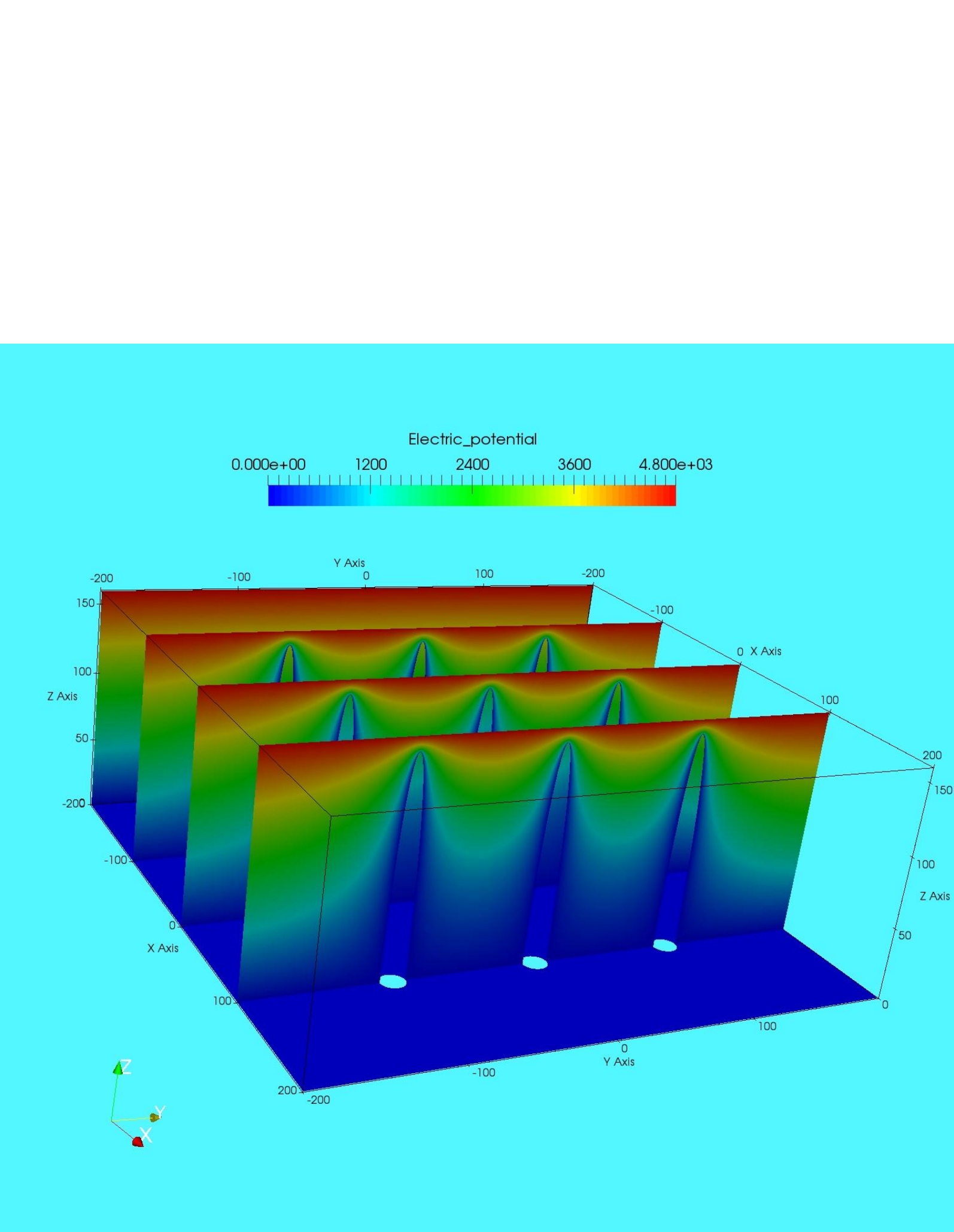}
\caption{A typical potential profile for a 9-emitter arrangement.}
\label{fig:potential}
\end{center}
\end{figure}

Multiple emitters provide a comprehensive testing ground for the effectiveness of the hybrid model. Even a $3 \times 3$ emitter
array can highlight the role of shielding and inverse shielding mediated through the anode. Consider thus, 9 identical
hemiellipsoid emitters arranged on a square grid with the nearest neighbour spacing denoted by $c$. Their height
$h  = 150\mu$m while $R_a = 500$nm and $D = 160\mu$m as in the previous section. The values of $\alpha_1$ and $\alpha_2$
are thus unchanged. The presence of other emitters leads to non-zero values of $\alpha_{S_i}$ and $\alpha_{{SA}_i}$
and this leads to a change in apex electric field. We summarize the results for $c = 2h/3$ in Table~\ref{tab:N9_1}.
Note that due to the symmetry, there are only 3 kinds of emitters. The one at the centre, the 4 nearest neighbours
and the 4 next-nearest neighbours located at the corners. Thus, analytical results show identical values
for the groups of 4 emitters. Note that, $\int E.dl$ computation uses grid data for both COMSOL
and analytical fields. A typical potential profile is shown in Fig.~\ref{fig:potential}.

\begin{table}[htb]
  \begin{center}
    \vskip -0.25 in
    \caption{Characteristics for a 9-emitter system with $c = 2h/3$ with
      $h = 150\mu$m.
      The symbols are CMSL (COMSOL), ANLY (Analytical), I (Current)
      and G (Grid).
      The correct value of $\int E.dl$ is 4800V. }
   \vskip 0.05 in
    \label{tab:N9_1}
    \begin{tabular}{|c|c|c|c|c|c|c|}
      \hline
      Location & $E_a$  & $E_a$  & $I_i$   & $I_i$ & $\int E.dl$   & $\int E.dl$   \\ 
      $x_i,y_i$ & CMSL & ANLY & CMSL &  ANLY & CMSL & ANLY (G) \\
      ($\mu$m) & (V/nm) & (V/nm) & (mA) & (mA) & (V) & (V) \\ \hline \hline
      0,0 & 4.595 & 4.498 & 0.128 & 0.0898 & 4792  & 4237 \\ \hline
      100,0 & 4.621 & 4.554  & 0.140 & 0.110  & 4792 & 4267  \\ \hline
      0,100 & 4.623 & 4.554 & 0.141 & 0.110 & 4794  & 4267 \\ \hline
      -100,0 & 4.632 & 4.554 & 0.146 & 0.110   & 4794 & 4267 \\ \hline
      0,-100 & 4.623 & 4.554 & 0.141 & 0.110 & 4793 & 4267 \\ \hline
      100,100 & 4.635 & 4.601 & 0.147 & 0.131 & 4795 & 4293 \\ \hline
      100,-100 & 4.632 & 4.601 & 0.150 & 0.131 & 4794 & 4293 \\ \hline
      -100,100 & 4.627 & 4.601 & 0.143 & 0.131 & 4796 & 4293 \\ \hline
      -100,-100 & 4.637 & 4.601 & 0.148 & 0.131 & 4793 & 4293 \\ \hline
    \end{tabular}
\end{center}
\end{table}

The average error in net emission current from the 9-pin array is about 18\% at $c = 2h/3$. As explained earlier,
this is a small error given the large deviations that small uncertainties in geometries can cause. The average
deviation of $\int E.dl$ using non-grid fields is 4.6\% while the error on using grid-data is higher at 10.9\%.
The COMSOL grid data is close to the true value of 4800V. The average transverse emittance $\sigma_x$  at the
anode plate ($D = 160\mu$m) using COMSOL data is 6.56nm while the average $\sigma_x$ using analytical grid data
is 9.92nm. The average $\sigma_x$ using non-grid analytical fields is expectedly lesser.

A similar study for $c = h/2$ leads to larger errors since the method under-predicts the apex field. Thus, the
net current is smaller by about 48\%, the error in non-grid $\int E.dl$ is 6.2\% while the error on using
grid-data is 12.5\%. The average $\sigma_x$ using COMSOL grid-data is 6.4nm while the average $\sigma_x$ using
analytical fields on the grid is 8.6nm.

It is expected that at a fixed anode-cathode distance $D$, the errors will further increase as $c$ decreases.
If, on the other hand, the emitters are randomly distributed having a mean separation $c$, the errors
are expected to be smaller as compared to the regular arrangement. This expectation stems from the observation
that pins that are closer to each other will have larger individual errors but their weighted contribution to
the net current is smaller due to the larger shielding effects. On the other hand, the sparsely populated regions
are expected to have smaller individual errors and larger weights. 

As the anode moves further away from the emitter tips, the effect of the term $\alpha_{{SA}_i}$ will be smaller
and shielding will dominate. Thus, the optimum emitter separation $c$ is expected to be larger than the
height $c$ in order to achieve maximum current density. In such a scenario, the errors on using the analytical
fields will be small.

\section{Concluding remarks}

The semi-analytical fields provide a useful, reasonably accurate and fast alternative for
a LAFE-based diode simulation. It has been tested here using a $3 \times 3$ grid and the results
are encouraging when the spacing is not too small compared to the height of the emitters. The
results can thus be used to simulate larger $N \times N$ grids (e.g. $N > 100$) or emitters placed
randomly. As mentioned earlier, this is still an idealization of a typical realistic situation
where emitter heights need not be identical or their apex radius of curvature may differ. These
aspects are of immediate interest and under investigation.
Other issues relate to space-charge effects and emitter degradation (in a statistical sense), both
of which can be approximately incorporated within the model proposed here.

Finally, the anode at close proximity is of relevance for gated diodes.
The results can be used to transmit electrons
through the anode plate with a pre-defined energy dependent transparency. Alternately, the gate can be simulated
using a circular apertures centred at each emitter. A COMSOL study using a 9-pin array shows that if the gate
aperture radius $R_g = h/6$ and $c = h/2$, the apex fields reduce as compared to a plate anode and are somewhat smaller
than the analytical values. Thus, the analytical modelling discussed in this paper approximates gated
diodes with circular apertures reasonably well and can be used to obtain estimates of the current and transverse
emittance.

\section{Author Declarations}

\subsection{Conflict of interest} There is no conflict of interest to disclose.
\subsection{Data Availability} The data that supports the findings of this study are available within the article.


\end{document}